# Optical Manipulation of Whispering Gallery Mode Microlasers for Precision Controlled Cellular Delivery

*Soraya Caixeiro[1,2*], Paloma Rodríguez-Sevilla[3], Kishan Dholakia[4,5,6], and Malte C Gather[1,4,7*]*

1 Humboldt Centre for Nano- and Biophotonics, Department of Chemistry, University of Cologne, Germany.

2 Centre for Photonics and Photonic Materials, Department of Physics University of Bath Claverton Down BA2 7AY, UK.

3 Nanomaterials for Bioimaging Group (NanoBIG), Departamento de Física de Materiales, Universidad Autónoma de Madrid, C/ Francisco Tomás y Valiente 7, 28049 Madrid, Spain

4 Centre of Biophotonics, SUPA School of Physics & Astronomy, University of St Andrews, North Haugh, St Andrews, UK.

5 School of Biological Sciences, University of Adelaide, Adelaide, South Australia, Australia

6 Centre of Light for Life, University of Adelaide, Adelaide, South Australia, Australia

7 Cologne Excellence Cluster on Cellular Stress Responses in Aging-Associated Disease (CECAD), University of Cologne, Cologne, Germany

* Correspondence to: scc201@bath.ac.uk; malte.gather@uni-koeln.de

**Abstract**

**Whispering gallery mode microlasers are known for their high Q-factors, characteristic emission spectra and sensitivity to local refractive index changes. This sensitivity combined with the ability of various cell types to internalise these microlasers provide unique opportunities for advanced biological studies, e.g. in single-cell tracking and intracellular sensing. Despite many advancements, achieving precise delivery of lasers to cells remains challenging, with traditional methods, such as microinjection, often also resulting in cellular damage. Here, we show that optical trapping is a promising solution for microlaser manipulation and therefore for their controlled and non-invasive delivery to target cells. By integrating optical trapping with microlaser-based refractive index sensing, we study the dynamics of microlaser-uptake by cells. We also find that in some cases the peaks in the emission spectra of our microlasers broaden, split, and shift, and we assign this to local inhomogeneities in refractive index surrounding the microlaser. This shows how optical trapping can further expand the unique toolkit offered by biointegrated whispering gallery mode microlasers.**

## Introduction

Whispering gallery mode (WGM) lasers, characterised by their high brightness and high Q-factor, have emerged as powerful platforms for biosensing[1]. These resonators exhibit remarkable sensitivity to changes in external refractive index and their small footprint allows seamless integration into living biological samples[2–5], opening up new possibilities for interrogation methods such as antigen binding sensing[6,7] and simultaneous tagging and tracking of single cells[8,9]. Despite the advancements in the field of biointegrated microlasers, there is a crucial need to enhance the precision delivery of these lasers at the single-cell level and enhance the efficiency of sensing and tracking while minimizing damage during cellular and tissue interactions.

Traditional bulk measurements of cells often fail to capture the nuanced variations at the single cell level. For example, changes in gene expression, metabolism, and response to stimuli are present among cells within the same tissue or organism. While various methods have achieved targeted single-cell delivery of nano and micro-particles into cells, many of these approaches involve invasive techniques that disrupt the cell membrane, such as microinjection, optical-poration, or electro-poration. In addition, these methods frequently demand advanced technical skills and may result in irreversible cellular damage. Microlasers have been delivered in a number of ways: microinjections has been successfully used for zebrafish heart[2], rabbits eye[10] and intravenously via a mouse tail[8]. However, the most commonly used delivery method involves adding microlasers to the cell media followed by overnight incubation. This method works most efficiently for phagocytic cells. For non-phagocytic cells, applying a cell membrane penetrating coating enhances uptake efficiency[11–13]. Despite its widespread use, this approach offers limited control over targeted cell delivery.

Optical trapping represents a promising solution to the limitations of traditional delivery methods. Three dimensional optical tweezers were first presented in 1986 by Ashkin and colleagues. He demonstrated that a tightly focused laser beam can trap a single microparticle in three dimensions[14]. Since then, optical tweezers have become a cornerstone tool for the life sciences[15,16], finding applications in various areas such as studies of single molecule biophysics[17], micromechanical sensing[18,19], cell surface imaging[20], and manipulation of microparticles in living organisms[21]. While micro-droplet lasers have been optically[22–24] and acoustically levitated [25], and WGM resonators have been trapped and used to spatially probe the refractive index of a fluid[26], the application of optical levitation for manipulation and biointegration of microlasers remains largely unexplored.

In this study, we present an optical tweezers platform tailored to facilitate the trapping and precise three-dimensional manipulation of a microlaser within a cellular environment. This platform enables the targeted and minimally invasive delivery of microlaser sensors to non-phagocytic cells. Subsequently, we can study and monitor of microlaser engulfment through refractive index sensing.

The marriage of optical trapping and microlaser sensing facilitates targeted and efficient delivery of microlaser sensors into cells, minimising damage, and elevating uptake statistics. In addition, it paves the way for a deeper understanding of the cellular uptake process of large particles, opening new opportunities in the understanding of cellular uptake processes. Understanding the intricacies of phagocytosis is pivotal not only in immunology but also in the broader context of drug delivery, and it has the potential to inform the development of novel therapeutic strategies with heightened specificity and efficiency[27].

**Results**

**Microlaser manipulation setup**

Figure 1a schematically illustrates the optical setup that allows microlaser trapping and simultaneous excitation. A long-pass dichroic beam-splitter combines a continuous wave laser with a wavelength of 1064 nm used for optical trapping and a pulsed solid-state laser with a wavelength of 473 nm (pulse width of <2 ns and repetition rate of 100 Hz) that is used as pump source for microlaser excitation. The microscope is configured to directly collect light onto a camera for cell visualisation (Fig. 1b) and a spectrometer for recording of the laser emission spectra (Fig. 1c). Further details of the experimental setup are available in the materials and methods section.

An example of the manipulation capabilities of our system is shown in Fig. 1b. Seven individual microlasers are strategically brought into contact with several HeLa cells, the human cell line used as model system in this study, demonstrating the ability of our platform to precisely position a set of microlasers and thus enhance their interaction with a biological sample.

The microspheres used in this study are fluorescently labelled polystyrene microspheres of 10.7±1.6µm diameter. They contain a highly fluorescent green-emitting dye that serves as gain medium and facilitates laser emission upon pulsed optical excitation with blue light and sufficient pulse energy to reach the laser threshold. In a previous study, the mean threshold of the fluorescent microspheres used here (i.e., the fluence above which they lase in aqueous media) was found to be 130 µJ cm$^{-2}$ ± 40 µJ cm$^{-2}$ [7]. Above the lasing threshold, a series of sharp peaks appears in the emission spectrum of the particles that dominate over the fluorescent background (Fig. 1c). These peaks are associated with sequential transverse magnetic (TM) and transverse electric (TE) modes, and their spectral position depends on the size of the resonator (microsphere) and the effective refractive index of the respective mode, also called the modal index, which in turn is related to mode number, polarisation (i.e., if TE or TM mode) and, importantly, the local refractive index of the medium surrounding the microsphere.

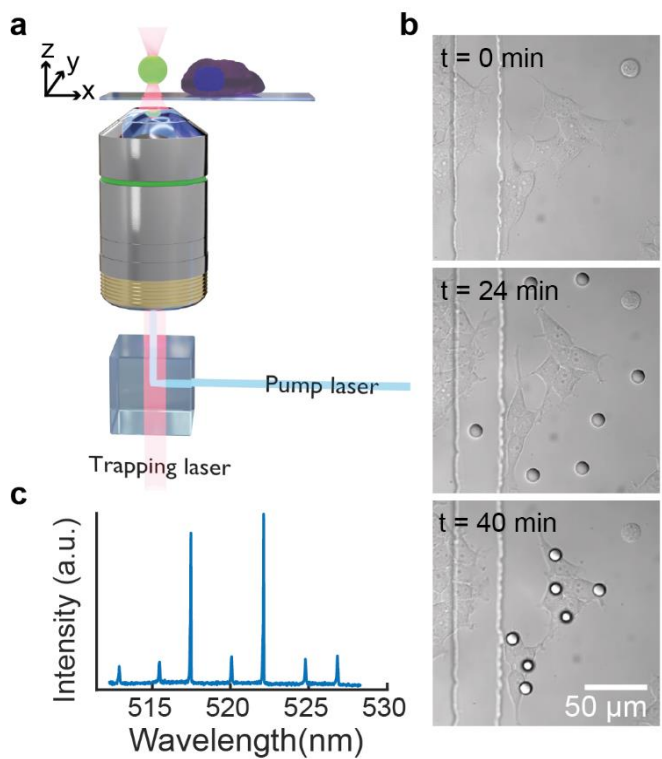

*Fig. 1* **Microlaser excitation and manipulation.** a) Schematic depiction of the setup used for microlaser trapping and simultaneous optical excitation on a cell-containing slide. b) Differential interference contrast (DIC) microscopy images illustrating the manipulation of 7 microlasers (circular objects) and their positioning atop a cluster of HeLa cells. c) Representative emission spectrum from an optically trapped microlaser pumped above the lasing threshold.

By precisely determining the wavelengths of at least two pairs of TE and TM lasing modes using a fit to a Voigt function, and subsequent fitting of these wavelength positions to an optical model, we independently determined the diameter of each microsphere and the external refractive index as detailed in Refs [7,9]. We previously reported the limit-of-detection for intracellular sensing with microlasers akin to those employed in this study to be 5.5 × 10$^{-5}$ refractive index units (RIU)[2]. At this resolution, WGM microlasers can be used as precise sensors of local refractive index, with direct applications in biosensing[2,9].

**Refractive index of microlasers in different media**

The intracellular refractive index is a key biophysical parameter and governs the propagation of light in the cell. Within the cell, different organelles exhibit different refractive indices. For instance, the cell cytosol has been documented to have a refractive index in the range of 1.36 to 1.39 RIU, while the nucleus exhibits values in the range of 1.355 to 1.365 RIU[28]. While these values offer indicative ranges, factors such as cell type and the specific phase of the cell cycle[29] can introduce variations. Seeking to utilise microlasers as reporters for inter- and intracellular refractive index, we conducted experiments under four distinct conditions to establish typical refractive index ranges for various microenvironments (Figure 2a). Microlasers were first immersed in deionized water (DIW) and cell media (DMEM, detailed composition, see materials and methods section) at room temperature, yielding a refractive index of 1.334 RIU ± 1 × $10^{-3}$ and 1.3401 ± 7 × $10^{-4}$ RIU, respectively (mean ± SD). This aligns with the literature value of refractive index for DIW and underscores that the presence of proteins and ions in DMEM results in an increased refractive index as expected[28,29].

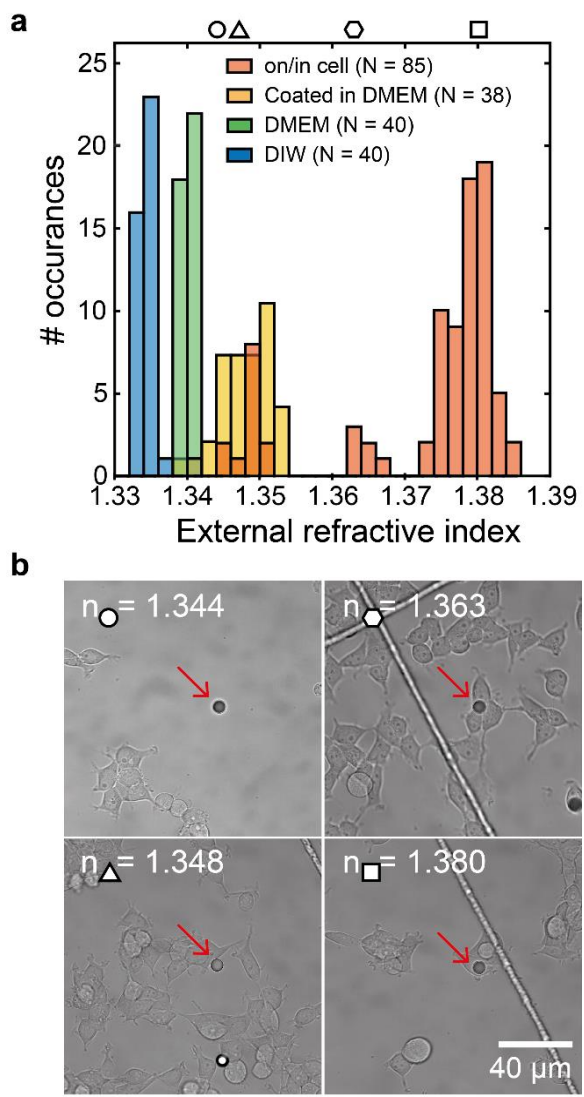

*Fig. 2 **Refractive index distribution in the surrounding of microlasers measured by evaluating their emission spectra.** a) Statistical analysis of refractive index for microlasers located in deionized water (DIW, blue), cell medium (DMEM, green), as well as for laser particles in DMEM and coated with lipofectamine (yellow) and in contact with or inside cells (red). b) Microscopy images of representative coated microlasers in cell media and corresponding fitted refractive index. Symbols correspond to histogram bars of equivalent refractive index in a.*

Non-phagocytic cells are capable of phagocytosing polymer microparticles[12], including microlasers exceeding 20 µm in diameter[11]—larger than those used in this study. The efficiency of this uptake can be significantly enhanced by coating the microlasers with lipofectamine[11]. In our study, we prepared the microlasers for cell integration by adapting this coating technique, as detailed in the materials and methods section. HeLa cells were cultured in a dish, and lipofectamine-coated microlasers were added in approximately a 1:2 excess. Following a 12 h incubation period, we determined the refractive index in the vicinity of both isolated microlasers and those in contact with cells (Fig. 2a and 2b).

The lipofectamine coating itself led to an increase in average refractive index to 1.348 RIU and a broadening in the spread of refractive indices (standard deviation) to $3 \times 10^{-3}$ RIU. This distribution serves as a baseline and reflects contributions from both the cell media and the microlaser coating. Next, we analysed spectra from microlasers in contact with or engulfed by cells. Within this dataset, three distinct distributions of refractive indices emerged. At the lower end of the distribution, an overlap with the refractive index values of isolated microlasers was observed. This corresponds to cases where there is little contact between the microlaser and the cell surface as illustrated in the $n$ = 1.348 subpanel of Fig. 2b (triangle). Here, the microlaser is positioned near and to the side of the cell and there is no evidence of membrane remodelling. The second distribution, and the least frequently observed refractive index values, is centred around 1.365 RIU. An example of a microlaser with an external refractive index in this range is depicted in Fig. 2b, $n$ = 1.363 RIU (hexagon). The microlaser is visibly between 3 cells and thus unlikely to be internalized by any cell; we identify this as a state where the microlaser is on top of cells and partially surrounded by cell membrane, causing a moderate increase in refractive index. Finally, the higher end of the refractive index distribution corresponds to cases where lasers exhibit extensive interaction with the cell membrane or have undergone some level of phagocytosis. The $n$ = 1.380 subpanel of Fig. 2b (square) shows a representative example of this. The higher refractive index values found in these cases are consistent with refractive index measurements of cells obtained via tomographic phase microscopy[30]. Additionally, the shape of the cell boundary follows that of the microlaser, indicating membrane remodelling around the microlaser and thus either advanced phagocytosis or complete microlaser engulfment.

We conclude that that there are two ranges of refractive indices relevant for assessing contact between microlasers and cells: a refractive index range falling between 1.337 and 1.359 is indicative of minimal or no contact between the microlaser and the cell body, conversely, a refractive index in the range of 1.360 and above suggest substantial contact, with the higher end indicating full uptake of the microlaser. We thus conclude that determination of refractive index can be a convenient and robust tool for assessing internalization of lasers, that can potentially be used in lieu of image analysis, especially in cases where accurate visual assessment of laser internalization is challenging.

**Refractive index sensing during manipulation of microlasers on cells**

To showcase the adaptability and efficacy of the microlaser platform in exploring the microenvironment around it, we investigate the change in local refractive index when microlasers are positioned on cells and other surfaces. In this experiment, we opt not to coat the microlasers with lipofectamine to minimise the adhesion to both the cell surface and the substrate. The microlasers used in this experiment have carboxylate groups on their surface, making them negatively charged and causing repulsion from the cell membrane, thus ensuring the microlaser does not adhere to surfaces easily.

Our initial focus is on evaluating the indentation dynamics of microlasers pushed into the cell surface. Microlasers are trapped above a single HeLa cell (Figure 3a), and the z position of the trap was incrementally lowered in 1-µm steps while simultaneously collecting spectra at each position. The relative movement was calibrated so that the lowest focal plane coincided with the point when the cell edges located near the substrate came into focus. As the focus is lowered and the optically trapped microlaser approaches the cell surface, it gently presses against the cell, inducing indentation, as illustrated by the schematic in Fig. 3a. At a certain point, the microlaser ceases to indent further despite the continued decrease in z height of the trapping beam and the causing the microlaser to appear out of focus. This indicates that the microlaser is making contact with the cell surface and is no longer moving with the focus of the trap. The external refractive index of the microlaser at each relative height step, as measured by fitting the position of the lasing peaks to our optical model, is shown in Fig. 3b. The refractive index is lowest at the highest point above the substrate and gradually increases as the height decreases and the microlaser comes in contact with the cell surface.

Careful analysis of the laser spectra recorded during the lowing of the laser onto a cell reveals that the individual peaks in the spectra broaden and, in some cases, split (Fig. 3c). This observation has been noted in previous studies[31]. We attribute this phenomenon to a lifting of the mode degeneracy between different WGM modes when the local refractive index in the vicinity of the laser becomes inhomogeneous. The modes generating laser emission are primarily localized in a plane perpendicular to the substrate because our optical pumping geometry is more efficient for these. However, multiple modes are likely excited in parallel. In cases where there is an inhomogeneous refractive index, such as when the laser comes into contact with the cell surface, the modes will 'see' varying effective refractive indices, consequently leading to slight differences in their resonant wavelengths. This then leads to a broadening or even splitting of peaks in the spectrum.

As the broadened and split peaks deviate from the typical Lorentzian shape, they cannot be accurately fitted with a single Voigt function. Instead, an algorithm was devised to identify the split peaks according to the overall peak width and to fit them to the sum of two Voigt functions (see materials and methods section). From the split peak positions, we then fit two refractive indices. The leftmost peaks in Fig. 3c (yellow) represent the lower end of the refractive index measured by the microlaser, while the rightmost peaks (green) represent the higher end of the refractive index range. This yields two refractive indices as also shown by filled circles in Fig. 3b and demonstrates how a single laser can sense the range of refractive indices present in its local environment. In reality, a single microlaser may support more than two degenerate modes at any given time; however, our findings suggest that due to resolution constraints on our current system, only the two extreme modes can be captured and consequently only the extremes of refractive index and the mean refractive index are determined with confidence.

Following the assessment of cell indentation effects, our investigation next extended to evaluating microlaser behaviour across various surfaces, particularly when in contact solely with the cell media. Fig. 3d shows a series of DIC microscopy images and schematic illustrations of the microlaser refractive index measurement while positioned on glass, on the cell surface and above these surfaces. The corresponding refractive index measurements for different positions are summarized in Fig. 3e, where the average refractive index corresponds to the unfilled circles. When in contact with the bottom glass of the Petri dish ($n_{\text{glass}} \approx 1.52$), the fitted external refractive index is higher than when the microlaser is elevated and thus completely surrounded by cell medium ($n_{\text{DMEM}} \approx 1.34$); this finding is consistent with literature results[32]. (Notably, the microlaser emission exhibits peak splitting in this example when the microlaser is optically levitating. This occurrence is exceptional and may potentially be attributed to adhered molecules or particles on the surface.) Next, we positioned the microlasers on top of different cells and placed them at various positions on the same cell, yielding a relatively wide range of peak splitting and fitted refractive index (*N*=7) represented in Fig 3e (blue region). This is reflective of the varying degrees of contact with the cell surface as well as the heterogeneity in the refractive index across the cell and from cell to cell. Subsequently, elevating the microlasers to a height where they remain in focus without touching the cell surface yields a decrease in refractive index (*N*=4), albeit not reaching the values attained for the laser hovering well above the glass surface. This last observation indicates that there may be some transfer of material from the cell to the surface of the microlaser upon cell contact, such as transmembrane proteins and lipids[33].

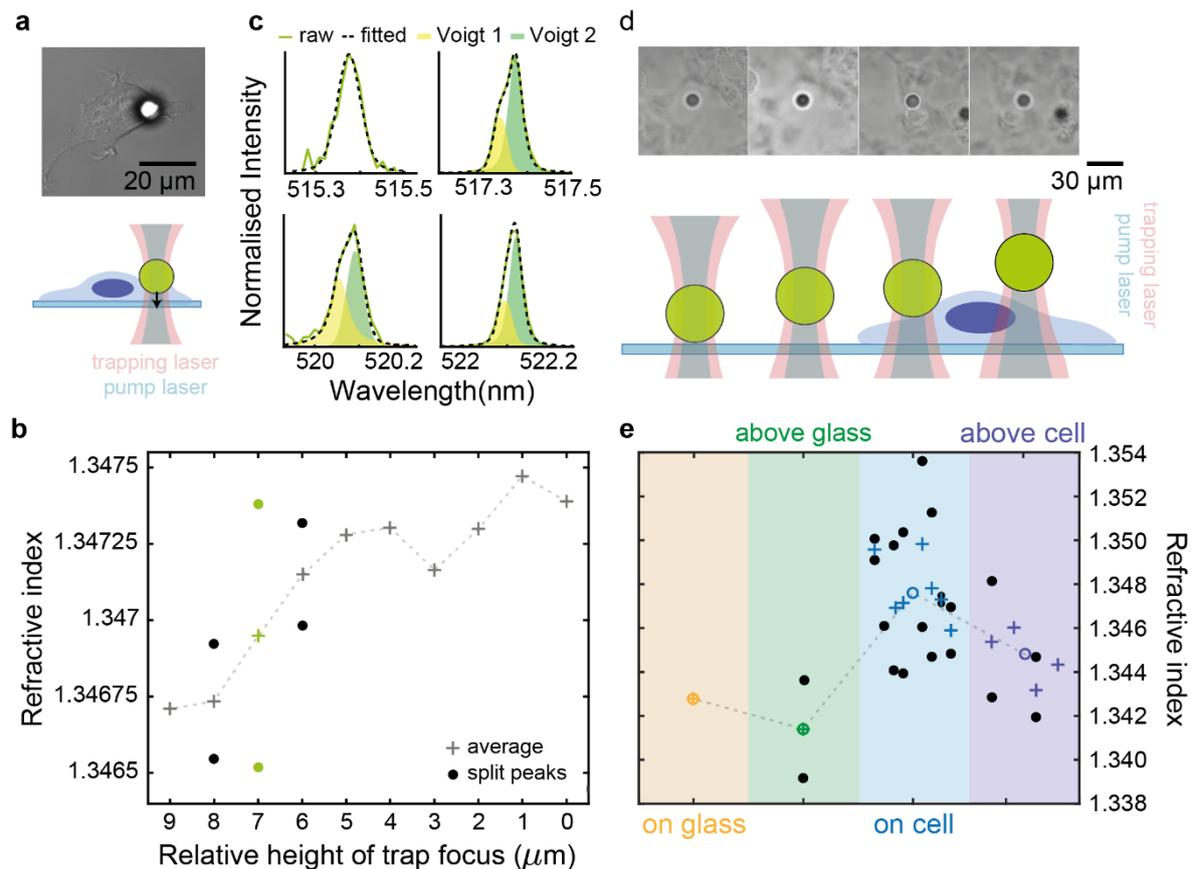

Fig. 3 **Microlaser manipulation on cells**. a) DIC microscopy image of microlaser indented into a cell by the optical trap and schematic illustration of the indentation process. b) Refractive index evolution while lowering the focus plane of the optical trap. Crosses indicate best fit for each position. Where peak splitting occurred, the filled dots indicate the refractive index fits based on the shortest and longest wavelength component of the split peaks. c) Representative peak splitting from a single microlaser during indentation, corresponding to green data points in b. d) Positioning of microlasers on glass, hoovering above the surface, in contact with cell surface, and hoovering over cells (from left to right). e) Refractive index measurements for the different positions indicated in d. There are N=1 measurements for the microlaser on glass (yellow)

*and above glass (green), N=7 measurements at different positions on top of the cell (blue), and N=4 measurements hovering above the cell (purple). Each cross symbolises the refractive index obtained for each spectrum, for spectra where peaks are split, the cross is the average refractive index obtained. The black circles represent the refractive index extremes from a spectrum with split peaks. The unfilled circles represent the overall average refractive index measured and are linked by a grey dash line.*

**Optical manipulation induced uptake**

Building upon our previous experiments investigating microlaser interactions and refractive index measurement, the next phase of our study explored how optical trapping and strategic positioning of microlasers can be exploited to enhance cellular uptake of microlasers. We compared two cell culture wells with HeLa cells; in one, microlasers were left undisturbed and therefore distributed randomly while in the other, approximately 200 microlasers that were initially not in contact with cells were put into contact with cells via optical manipulation, as illustrated in Figure 4a. Each microlaser was optically held on top of a cell for about 10-20 seconds and the focus lowered similarly to the procedure employed during the indentation experiments above. At this stage, the microlaser is attached to the cell surface, and we find that the stiffness of our optical trap was not sufficient to completely sever this attachment again.

In some instances, the formation of a flexible membrane tether was observed, allowing for stretching between the microlaser and the cell membrane. This phenomenon, documented in literature[33,34], serves as a valuable tool for probing membrane elasticity, surface tension, and bending modulus. Within this framework, the microlaser may become displaced from the cell surface, while the tether persists as a resilient link. When the optical trap is moved too far from the tethering point, the microlaser either follows the trap or recoils back. This behaviour indicates that binding between the cationic-lipid coating of the microlaser and the receptors on the cell surface is present, and that the uptake process has been initiated.

Following an incubation period of 12 hours, measurements of refractive index were conducted in each well to assess the number of internalized microlasers (Fig. 4b and c). To mitigate potential measurement bias, random squares in the gridded dish were selected for the measurements, and the refractive indices of all microlasers within each square were measured. In addition, the wells were measured in an alternating fashion, to remove any effect of a potential further increase in uptake over the cause of the measurements. (An alternative would have been to fix all cells prior to the refractive index measurement. However, this was not done as the chemicals used for fixation induce crosslinking of proteins within the cell[35], which can adversely affect cellular morphology and has been demonstrated to decrease the refractive index of both the cell and its organelles[36], thus potentially skewing the result of the internalization measurement.)

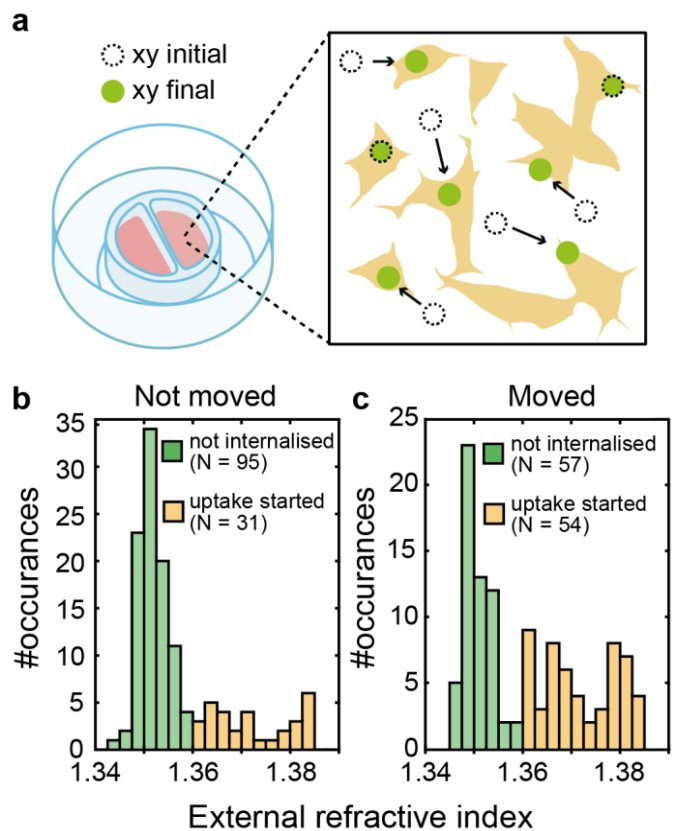

*Fig. 4 **Enhanced microlaser uptake through optical manipulation**. a) Schematic of manipulation experiment. b) Statistical analysis of refractive index distribution for N= 126 randomly placed microlasers. c) Statistical analysis of for N= 111 positioned on cells 12 h prior to the refractive index measurement using optical manipulation.*

As before, minimal contact between microlaser and cell was defined as refractive index lower than 1.360 RIU, while a refractive index higher than 1.360 RIU was taken as having good contact. Remarkably, nearly half of the microlasers assessed in the manipulated well exhibited a refractive index greater than 1.360 RIU. Conversely, in the control well, only a quarter of the microlasers demonstrated a notable interaction with the surrounding cells, which is similar to the fraction reported in the literature for similarly sized microlasers seeded on HeLa cells[11]. These findings suggest that displacing the microlasers can yield improved uptake and enhance the interaction between microlasers and cells. The exact ratio of internalized lasers will depend on the number of the initially seeded cells and the microlaser count. However, the trend observed here demonstrates a positive impact of optical manipulation on cellular interaction and uptake dynamics, which is likely to hold for other levels of laser and cell density.

**Microlaser uptake dynamics**

Lastly, we investigate the dynamics of microlaser uptake. For this, microlasers were again trapped and placed on a group of cells, using the method described in the previous experiment (Figure 5a). The microlasers were now observed continuously throughout the uptake process. The uptake is associated with a change in the height of the microlasers as seen by the lasers gradually coming into focus on the microscopy images of the cells[37]; a representative example of this is shown in the inset to Fig. 5a. Fig. 5b shows a schematic illustration of the uptake process and the associated change in height above the substrate. The plasma membrane undergoes deformation mediated by the actin cytoskeleton of the cell, which leads to the formation of a phagocytic cup, as well as protrusions and remodelling of the plasma membrane around the microlaser[38]. Transport within the cell commences and the focal plane of the microlaser lowers as time progresses until it ultimately approaches that of the cell.

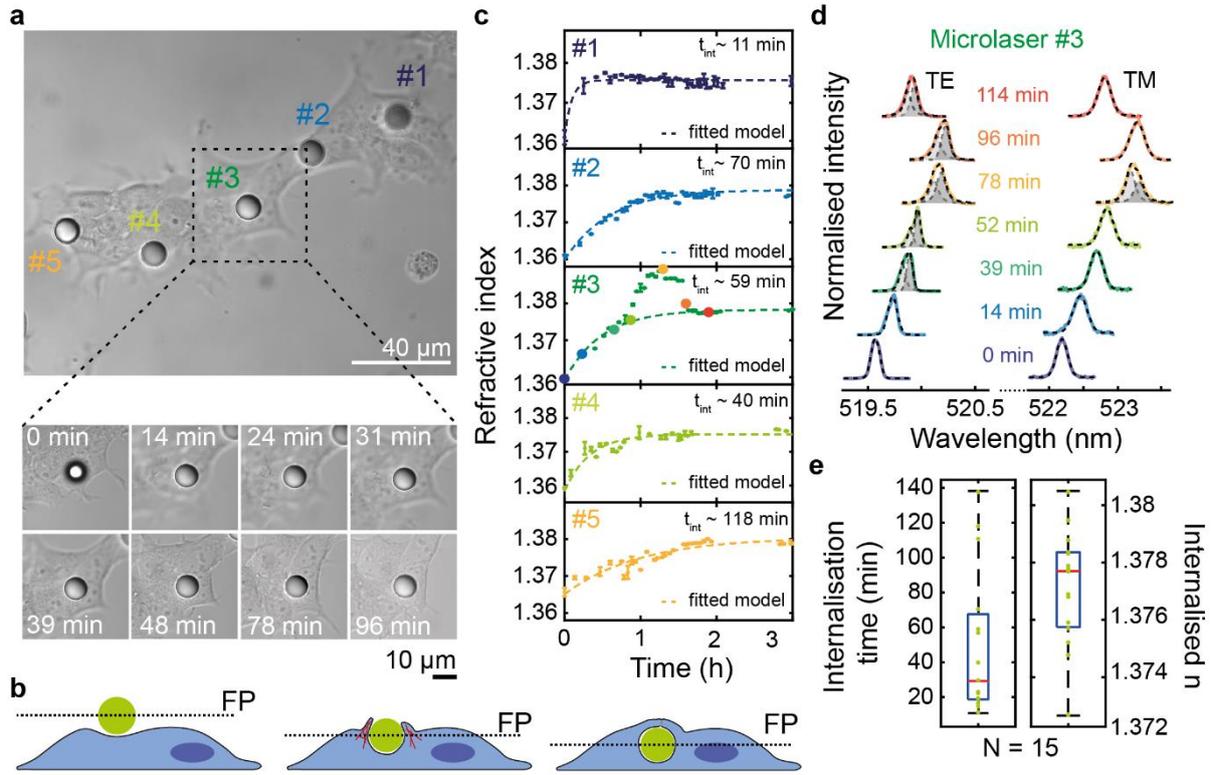

Fig. 5 **Uptake dynamics of microlasers.** a) Microscopy image of microlasers #1 - #5, each of which is placed onto a cell. Inset, representative timelapse showing how Microlaser #3 gradually moves downwards and into focus. b) Schematic representation illustrating the different steps of the microlaser uptake mechanism. c) Evolution of refractive index during the internalisation of microlasers #1-#5 (symbols) and nonlinear least square fit of the refractive index to equation 1. Error bars represent the standard deviation from the mean refractive index obtained from several spectra. d) Temporal evolution of a transverse electric (TE) and a transverse magnetic mode (TM) peak in the lasing spectrum of Microlaser #3, colour coded corresponding timepoints indicated in c. e) Internalisation statistics of 15 microlasers showing the time for internalisation to complete and the final refractive index. Error bars represent the full range of values, boxes the standard deviation and the red lines represent the median.

Fig. 5c tracks the dynamics of the uptake process for the five microlasers depicted in Fig. 5a over 3 hours by plotting the evolution of their external refractive index over time, starting from the placement of the microlasers on the respective cell. At least 3 spectra from each microlaser are acquired at each time point, fitted as described in the materials and methods section, and then plotted on the graphs with corresponding error bars.

The uptake traces are fitted with an internalisation model adapted from Ref. [39] (dashed lines in Fig. 5c). This model describes the progressive increase of the external refractive index over time and its subsequent stabilisation by the following equation:

$$n(t) = n_0 + \Delta n(1 - e^{-(t-t_0)/\frac{t_{int}}{2}}),$$

(1)

where $n_0$ is the initial refractive index, $\Delta n$ is the total refractive index increase, $t_{int}$ is the internalisation time and $t_0$ is the time at which internalisation starts. The five traces shown behave similarly whereby the refractive index increases steadily from an initial value ranging from 1.36-1.365 until it plateaus at around 1.37-1.38. This is thought to be the point where the internalisation is complete, and the membrane fully ingulfs the microlasers. The TE and TM peaks from Microlaser #3 are depicted in Fig. 5d. Each pair of peaks has a corresponding coloured symbol on the uptake graph (Fig. 5c). The peak positions shift in correspondence with the uptake graph, where an increase or decrease in the refractive index corresponds to a red shift or blue shift, respectively. At first, both peaks red shift showing no peak splitting, but as time progresses and the refractive index increase, the peaks split and exhibit spectral broadening due to the inhomogeneous distribution of the refractive index that is evanescently coupled to the resonant modes. While small variations in refractive index are accredited to microlaser transport within the cell, the refractive index evolution of Microlaser #3 clearly deviates from the general internalisation trend, as its refractive index first increases to a value well above 1.38 and then decreases again before reaching a plateau There are several plausible explanations for this: Firstly, it is possible that the Microlaser #3 is in close proximity to a cellular organelle with higher refractive index, such as mitochondria or that Microlaser #3 is situated within a higher refractive index region of the cytosol, which has been reported to have a refractive index of up to 1.39[28]. This interpretation is supported by the evolution of the TE and TM peaks from Microlaser #3 (Fig. 5d). As observed before, some peaks exhibit spectral broadening, in particular beyond the 52 min time point, when the refractive index increases to a value above the fit line in Fig. 5c. As argued before, this splitting is most likely due to an inhomogeneous spatial distribution of the refractive index during microlaser internalization. Another explanation for the strong increase and subsequent decay in external refractive index would be that a phagocytic cup is formed around the surface of the microlaser and a process of exocytosis, whereby the cell expels the content of the phagosome, has begun; however, a more significant reduction of the refractive index would be expected for this latter case.

The internalization model is fitted to the refractive index evolution of $N = 15$ microlasers, for which internalisation was deemed successful in order to obtain the average internalisation time $t_{\text{int}}$ and the final internalised refractive index $n_0 + \Delta n$ (Fig. 5e). The success of internalisation is assessed based on two criteria: the alignment of the focal plane of the microlaser with either the nucleus or the cell edge, and the observed increase in refractive index. (Instances of unsuccessful internalisations are compiled in Supplementary Figure S1).

We observe a significant variation in internalisation times, ranging from approximately 10 minutes to over 2 hours. Nevertheless, the majority of internalisation events occurred within the initial hour of placement of the microlaser on top of the cell, with a median uptake time of approximately 30 minutes.

The final fitted refractive index varies over a relatively narrow range, with values ranging between 1.373 and 1.381 RIU and a median of 1.378 RIU. This consistency of the refractive index values demonstrates that the spectra of a microlaser can be utilised as a reliable indicator of microparticle internalisation.

**Discussion**:

Our study showcases the successful trapping and precise positioning of microlasers in contact with cells. Not only can microlasers be placed on cells to promote internalisation and indent the cell surface, but they can also detect the refractive index of their surrounding providing insights into their local optical environment. This novel synergy between microlasers and optical manipulation opens up avenues for integration with other biomechanical and microrheology studies[19], expanding our understanding of cell mechanics, and the uptake mechanism of non-phagocytic cells.

The ability to deliver microlasers to individual cells and enhance uptake efficiency presents a minimally invasive and highly precise method for microlaser delivery, offering cell and cell-type selectivity without having to resort to separate microlaser incubation with each cell type and/or cell sorting to select internalised lasers[40]. Furthermore, the refractive index monitoring sheds light on the intricate interplay between microlasers and cellular environments during engulfment. We have documented distinct optical responses including peak splitting and spectral broadening, and we established their correlation with factors such as the refractive index distribution and the degree of cellular engulfment. Notably, the sensitivity of microlasers to changes in refractive index highlights their potential as highly responsive optical probes for studying cellular dynamics.

We observed a wide range in microlaser internalisation times, spanning from 10 minutes up to 2 hours. The phagocytosis process can be influenced by many factors including material stiffness, shape, size, and surface chemistry of the target to be uptaken[37,38,41] as well as the cell type and growth phase[2,42]. Notably, there are limited works on the dynamics of particle uptake, with the majority covering smaller beads sizes and using phagocytic cells such as macrophages[37]; this is likely due to experimental challenges, which our combined trapping/microlaser approach can address, at least in part. Our findings suggest longer uptake times, which we attribute to both the non-phagocytic nature of the cell line used and the relatively large size of the microlasers employed compared to literature. The phagocytic cup growth is proportional to the square root of the uptake time[43], in other words and perhaps not surprisingly, larger particles require more time to forming the phagocytic cup, which is essential for particle engulfment.

There is evidence that the external refractive index measured by the microlaser upon initial contact with the cell is a good indication of future uptake, with high index indicating a larger chance of uptake and a shorter uptake time. This could suggest that cell indentation aids the formation of the phagocytic cup. However, further experiments are required to fully validate this observation.

The methodology used for cell indentation in this study draws parallels with established approaches used to investigate the mechanical properties of cells, including their stiffness and elasticity[44]. The measurement of refractive index thus might lead to insights into the quantification of indentation, membrane remodelling as well as the dynamics of phagocytic engulfment[38]. This future avenue of inquiry can complement current techniques and further our comprehension of cellular mechanics and structural dynamics.

To extend the scope of the work presented here, holographic optical tweezers[45] might be used to create individual simultaneously controlled optical traps for parallel manipulation of multiple microlasers. Equally employing hyperspectral strategies —such as push-broom spectroscopy[46], widefield hyperspectral imaging[47], and hyperspectral confocal imaging[9]— could complement this and yield larger sample sizes and thus improved statistics. Moreover, achieving more selective excitation of degenerate WGMs could provide insights into the 3D environment surrounding the microlaser and thus the cell. Recently, the use of optically manipulated micromirrors [32] and light sheet microscopy configurations[48] has allowed for selective mode excitation and these approaches could also be applied here.

Another potential advancement is to transition to smaller lasers such as nanowires[4] and disk-shaped nano-lasers[5,7], which present similar sensing capabilities, with lower lasing thresholds and a reduced footprint. The reduced size of nano-disk lasers and nanowires could enable more precise sensing within cells and tissues. This enhanced precision could lead to improved spatial resolution and sensitivity in biosensing applications, facilitating the study of cellular dynamics and interactions at the nanoscale. However, it is well documented in literature that rigid spherical objects have better uptake outcomes, and wire- and disk-shaped objects are harder for cells to wrap around, especially when these objects are lying flat on the cell. On the other hand, if positioned perpendicular to the cell surface, tip-first, the uptake can be significantly improved[41]. Trapping mechanisms can aid in achieving this orientation.

The larger refractive index of nanowires and disk-shaped nano-lasers relative to the silica spheres used widely in optical trapping poses a challenge for three-dimensional confinement in a conventional gravitational trap, such as the one used in this study. This is due to significant destabilising scattering forces, due to their size, high refractive index, and non-spherical shape[49]. To address this issue, more complex trap geometries like counterpropagating traps and wavefront shaping may be necessary. Such advancements would not only enable optical manipulation of smaller, non-spherical lasers but also offer control over their orientation within the trap and on the cell surface.

Moving forward, understanding the nuanced behaviours of microlasers in biological contexts holds significant implications for a wide range of applications, from biosensing to targeted drug delivery. By unravelling the intricacies of microlaser-cell interactions, we can harness their unique optical properties to advance our understanding of cellular processes and potentially pave the way for novel biomedical technologies with enhanced precision and sensitivity.

**Materials and methods**

*Optical setup*

The optical trapping, pumping, and laser spectroscopy components were integrated into a standard inverted fluorescence microscope (Nikon, TE2000) with DIC imaging capabilities. The trapping laser was a solid-state continuous laser (Laser Quantum, Opus 1064) with up to 2 W power at 1064 nm. Spatial filtering ensured a Gaussian beam profile, and power at the sample was maintained between 50 mW and 250 mW, enough for stable optical manipulation of the microlasers. A Q-switched diode-pumped solid-state laser (Alphalas) with a wavelength of 473 nm, pulse width >2 ns, and repetition rate of 100 Hz, excited the microlasers, the pulse energy was controlled via neutral density filters. This laser was made colinear with the trapping laser using a dichroic beam splitter. Both beams, trapping and excitation beam, were expanded to overfill the microscope objective (Nikon CFI Plan Apo VC, NA 1.4) to achieve diffraction-limited spots. A broadband 50/50 beam splitter was employed to direct the beam into the objective.

The emission from the microlaser was collected by the same objective and passed to the camera port of the microscope. The image was relayed to a spectrometer (Andor Shamrock 500i with a 1200 lines per mm grating and Andor Newton DU970P-BVF) and imaged by a cooled sCMOS camera (Hamamatsu, Orca Flash 4.0v2) using a series of relay lenses and dichroic beam splitters. Transmitted red illumination is employed for simultaneous DIC imaging and microlaser excitation.

During lasing experiments, cells were maintained in a humidified stage-top incubator system (Okolab, H301) set to 37 °C and purged with a 5%:95% $CO_2$:air mixture.

*Cell culture*

HeLa cells were cultured in DMEM phenol free media (Thermofisher) supplemented with 10%v/v foetal bovine serum, 1%v/v penicillin-streptomycin in vented flasks at 37 °C and 5% $CO_2$. Cells were passaged when the culture reached ≈85% confluency. Cells were detached from the flask with room temperature TrypLE (Themofisher) for 2 minutes and centrifuged at 200 g for 5 min and resuspended in culture media and replated.

For the trapping experiments, cells were counted using a hemocytometer, and the appropriate number of cells as indicated by the dish and inset supplier (Ibidi GmbH) was added. Glass bottom dishes, both gridded and non-gridded, were utilized for the experiments.

*Microlaser preparation*

Microlasers (Polyscience) were suspended sterile DIW at a density ~0.025% w/v, equivalent to approximately $10^5$ microlasers. 1% v/v of Lipofectamine 3000 (Invitrogen) was added to the suspension, and the suspension was vortexed for 5 s and then shaken on an orbital shaker for 30 minutes at 300 rpm at room temperature. Subsequently, the suspension was centrifuged for 5 minutes at 12,000 x g, and the supernatant was removed. This process of centrifugation and resuspension was repeated three times, with the resuspending solution being replaced with cell media on the final cycle, ensuring an appropriate volume for adding the desired quantity to the dish or inset.

For the measurements of uptake statistics a gridded petri dish, featuring a silicon well insert to segregate two experimental conditions, was used. HeLa cells were evenly distributed in both wells at a density of $6.5 \times 10^4$ cells/cm$^2$ and allowed to settle overnight. Subsequently, the prepared microlasers were introduced simultaneously to both wells.

*Spectrum peak splitting fitting*

The splitting of modes is determined by comparing the width at 20% of the peak height to the undisturbed mode, i.e. before it is in contact with the cell. If the width is more than 1.5 times larger than the undisturbed mode, it is assumed to be split, and the fitting algorithm proceeds to fit a sum of two Voigt peaks. The minimum full width at half maximum (FWHM) is set to the resolution of the spectrometer, using the undisturbed FWHM as a seed value. Additionally, a minimum signal-to-noise ratio between the peak and the background noise is set such that very noisy peaks are not fitted with two Voigt peaks, as observed for the first peak of the spectra in Fig. 3c. This measure is required as fitting noisy peaks as split peaks yielded erratic results.

**Acknowledgements**

Authors wish to thank Prof. Marcel Schubert and Dr Graham D. Bruce for the fruitful discussions.

This work was financially supported by EPSRC (grant no. EP/P030017/1) and the Humboldt Foundation (Alexander von Humboldt Professorship to M.C.G.). P.R.-S. was supported by a Juan de la Cierva – Incorporación scholarship (Grant No. IJC2019-041915-I) and the Convocatoria del programa de ayudas UAM-Santander para la movilidad de jóvenes investigadores- 2021. KD acknowledges support of the Australian Research Council (through grants DP220102303 and FL FL210100099)